# On Intentional Attacks and Protections in Complex Communication Networks

Shi Xiao and Gaoxi Xiao*
Division of Communication Engineering
School of Electrical and Electronic Engineering
Nanyang Technological University, Singapore 639798
Email: shixiao@pmail.ntu.edu.sg, egxxiao@ntu.edu.sg

***Abstract*—Being motivated by recent developments in the theory of complex networks, we examine the robustness of communication networks under intentional attack that takes down network nodes in a decreasing order of their nodal degrees. In this paper, we study two different effects that have been largely missed in the existing results: (i) some communication networks, like Internet, are too large for anyone to have global information of their topologies, which makes the accurate intentional attack practically impossible; and (ii) most attacks in communication networks are propagated from one node to its neighborhood node(s), utilizing local network-topology information only. We show that incomplete global information has different impacts to the intentional attack in different circumstances, while local information-based attacks can be actually highly efficient. Such insights would be helpful for the future developments of efficient network attack/protection schemes.**

## I. INTRODUCTION

Recently there has been extensive research interest in studying large-scale real-world systems by formulating them into various network models. It is found that different systems, varying from Internet, world-wide web (WWW), airline transportation systems, food web, protein-protein reactions, to co-authorship, even terrorism organizations, when formulated into network models, sharing some stunning common features. Such observations give rise to a new area called *complex networks* [1, 2].

The most important and most popular complex network model is probably the *scale-free network* [1, 2]. In such networks, the number of nodes with a nodal degree $k$ is basically proportional to $k^{-\alpha}$. (For most real-world systems, the value of the exponent $\alpha$ lies between 2 and 3 [1], the reason of which, however, remains largely unknown.). It is claimed that all the systems we mentioned above can be formulated as scale-free networks [1, 2].

This work was partly supported by Microsoft Research Asia (MSR-A)
*Corresponding author (Phone: +65 6790 4552)

There have been debates on whether the communication systems, especially the Internet, are indeed scale-free networks [3]-[5]. While most statistic results are showing scale-free nodal-degree distributions in communication systems, the argument is that the traceroute method (the method of sending traceroute packets to record the IP addresses of all the routers they have passed through) which we are using to sample the Internet topology, leads to biased statistical results [5]. Though the debates may not be resolved until a better sampling method is devised, it seems to be agreeable that at least the tracerouted part of the Internet, the part that the starting-node set can steadily reach, usually the part that matters most to the starting-node set as well, indeed forms up a scale-free network.

Research on complex networks helps achieve better understanding of communication systems. One of the most important results reported is that the scale-free networks are highly robust to *random failures* (defined as network nodes failing one after another in a random sequence) yet fragile under *intentional attacks* (defined as network nodes being crashed one by one in a decreasing order of their nodal degrees) [6][1]. In the Internet, though large-degree hub nodes typically are on the network edge instead of network backbone [8, 9], the removal of the hubs will nevertheless leave a large number of nodes disconnected from the network. The consequence, therefore, can still be disastrous.

We are motivated to study the robustness of communication systems under intentional attacks from a complex network point of view. Specifically, we notice that in the existing results, to the best of our knowledge, two different effects have largely been ignored. First, complex communication systems such as Internet or WWW are so large, that no one really has the global nodal-degree information for executing an accurate intentional attack. Second, the real-world intentional attacks, if they can be executed at all, generally have to "propagate" through the network, attacking the neighborhood nodes adjacent to those "occupied" nodes step by step. And such attacks typically are based on local information of network topology only. In this paper, we call such kind of attacks the *distributed attacks*.

We examine the above two effects through extensive simulations. It is shown that (i) different missed knowledge of network topology has different impacts to the efficiency of intentional attacks. Hence hiding network-topology information may help strengthen network robustness under certain circumstances; (ii) on the other hand, distributed attacks with no global network-topology information can achieve very high efficiency, sometimes almost as efficient as the intentional attack based on accurate global information.

[1] Theoretical analysis on what kinds of networks are most tolerant to random failures and intentional attacks respectively can be found in [7]. As far as we know, however, there is no thorough discussion on whether these "optimal" networks can have wide real-world applications, and more importantly, whether they can emerge in self-organized extra-large systems.

Rest parts of this paper are organized as follows. In Section II, we study the efficiency of the intentional attack based on incomplete/inaccurate network topology information. A few distributed attack schemes are evaluated in Section III. Finally, Section IV concludes the paper.

## II. Intentional Attacks Based on Incomplete Network-Topology Information

The robustness of a network is typically measured by its *biggest cluster size* (i.e., the number of nodes in the biggest *connected* node set versus the number of nodes in the whole network) and the *cluster diameter* (i.e., the average length of the shortest paths between all the node pairs in the biggest connected node set) under attacks [1, 2, 6, 10]. If a node has been attacked, we call it *crashed*; otherwise, we call it a *live* node.

As mentioned before, the intentional attack is defined as the attack that brings down network nodes in a decreasing order of nodal degrees [6]. It is shown by extensive simulations [6] and theoretical analysis [10] that the scale-free networks are fragile under intentional attacks. To provide some examples, we simulate the intentional attacks in the well-known Barabási-Albert (BA) model with 10,000 nodes and 20,000 links [1, 6] as well as a real-world Internet model containing 6,470 nodes connected by 12,560 links [11]. In our simulations, we iteratively take down the largest-degree node in the *current* network until the network is broken into small pieces. It is shown in Fig. 1 that the BA model is crashed when roughly about 14% of all the network nodes are taken down, while the Internet model has an even lower "threshold of crash" at about 2.7%. For comparison purpose, we plot in Fig. 1 the network behavior under random failures. We see that both networks are highly tolerant to such failures.

To study the efficiency of intentional attack where accurate global information of network topology is *not* available, we consider two different cases as follows:

- Due to incomplete knowledge of network nodal-degree distribution, a big hub is missed in the intentional attack. As an example, we test the special case that only the biggest hub is missed while all the other hubs are accurately taken down;
- The big hubs are all taken down, yet some medium-sized hubs are missed in the attack. Specifically, in the BA model, we test the cases where the first 1% biggest hubs are all taken down and after that, 10% and 50% of the next top 3% hub nodes (with nodal degrees varying from 12 to 23) are missed respectively. The same percentages apply to the simulations in the Internet model, where the degrees of the missed medium-sized nodes vary from 10 to 40.

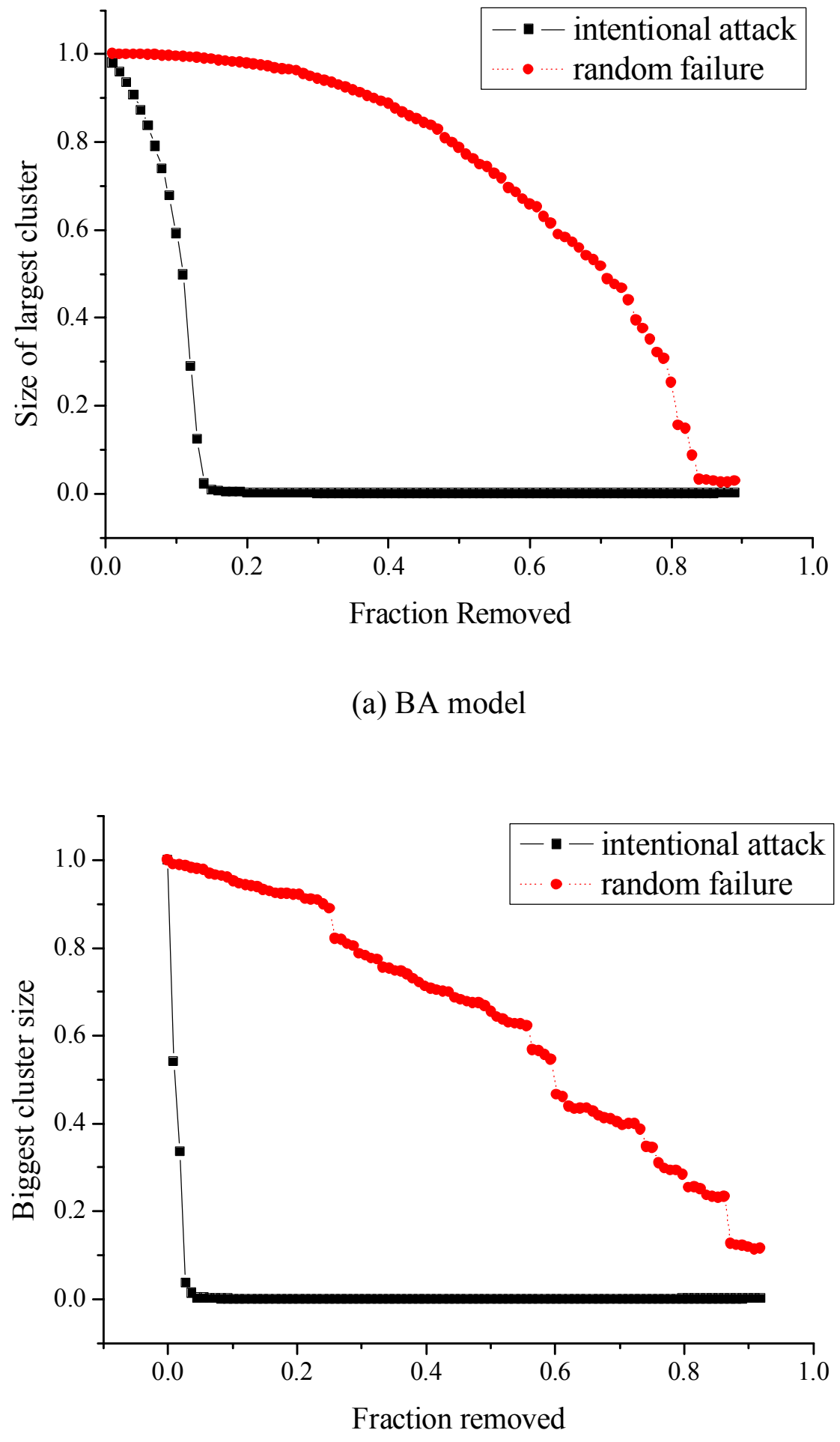

(a) BA model

(b) 6470-node Internet model

Fig. 1. Robustness of the scale-free networks under intentional attack and random failure, respectively.

The simulation results are presented in Fig. 2. We observe that missing the single biggest hub makes the intentional attack drastically less efficient. In the BA model, 21% of all the nodes have to be removed before the network goes crashed, denoting a 7% increase in the threshold of crash. In the Internet model, the effect is even more significant. The threshold of crash jumps from a very low 2.7% to a rather high 30%. The fragileness of the Internet model under intentional attack and this significant increase in the threshold of crash can be explained by the same reason: the Internet hubs are of very high degrees (The biggest hub has a degree of 1,458, while in the BA model it is only 386.). Therefore the survival of a single hub may drastically enhance network connectivity. The lesson we can learn from such observations is obvious: it is worth all the efforts for the offending (defending) side to take down (protect) the big hubs.

Missing a certain portion of the medium-sized nodes also degrades the efficiency of intentional attack, as we can observe in Fig. 2. Considering the fact that it is extremely difficult to identify all the medium-sized nodes in extra-large communication systems, we argue that the high possibility of missing some medium-sized nodes may indeed impose a threat to the efficiency of intentional attack. Furthermore, comparing the results of missing 10% and 50% medium-sized nodes respectively, we argue that, for the defending side, in certain circumstances it may be in fact a more effective strategy to protect/hide as many medium-size nodes as possible rather than protecting the obvious targets of big hubs.

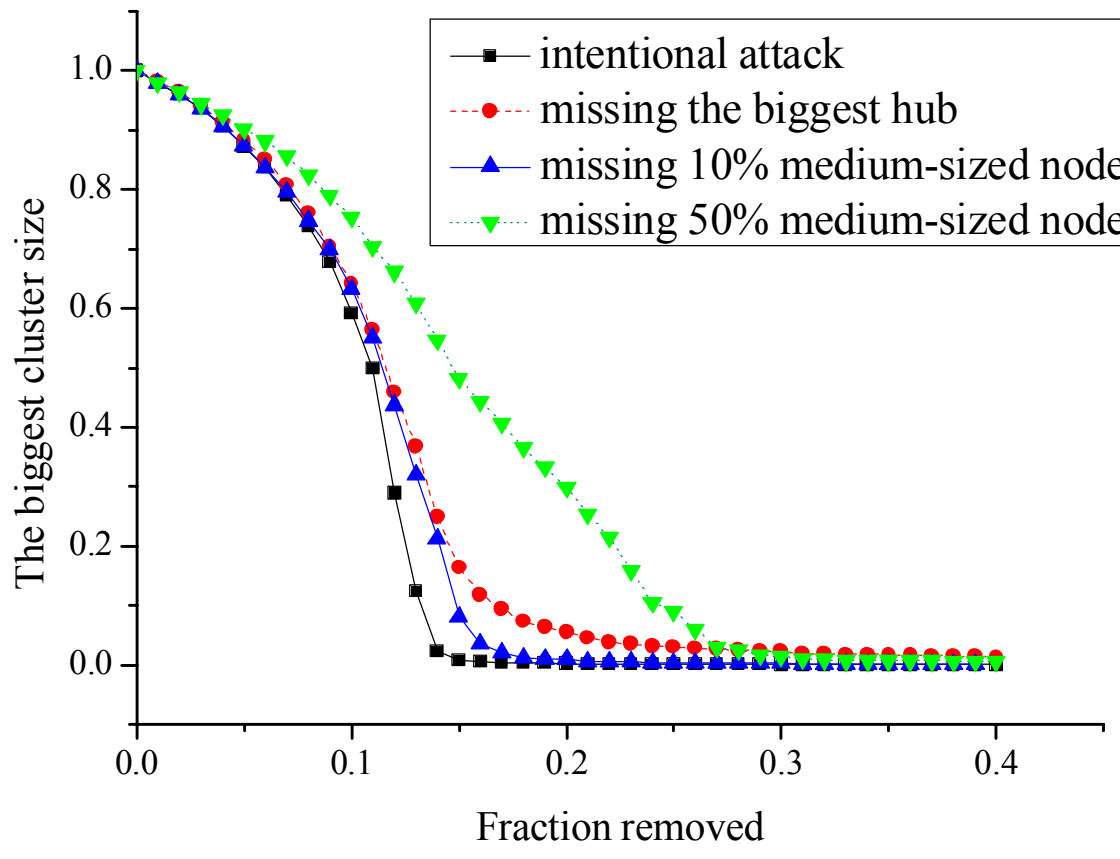

(a) BA model

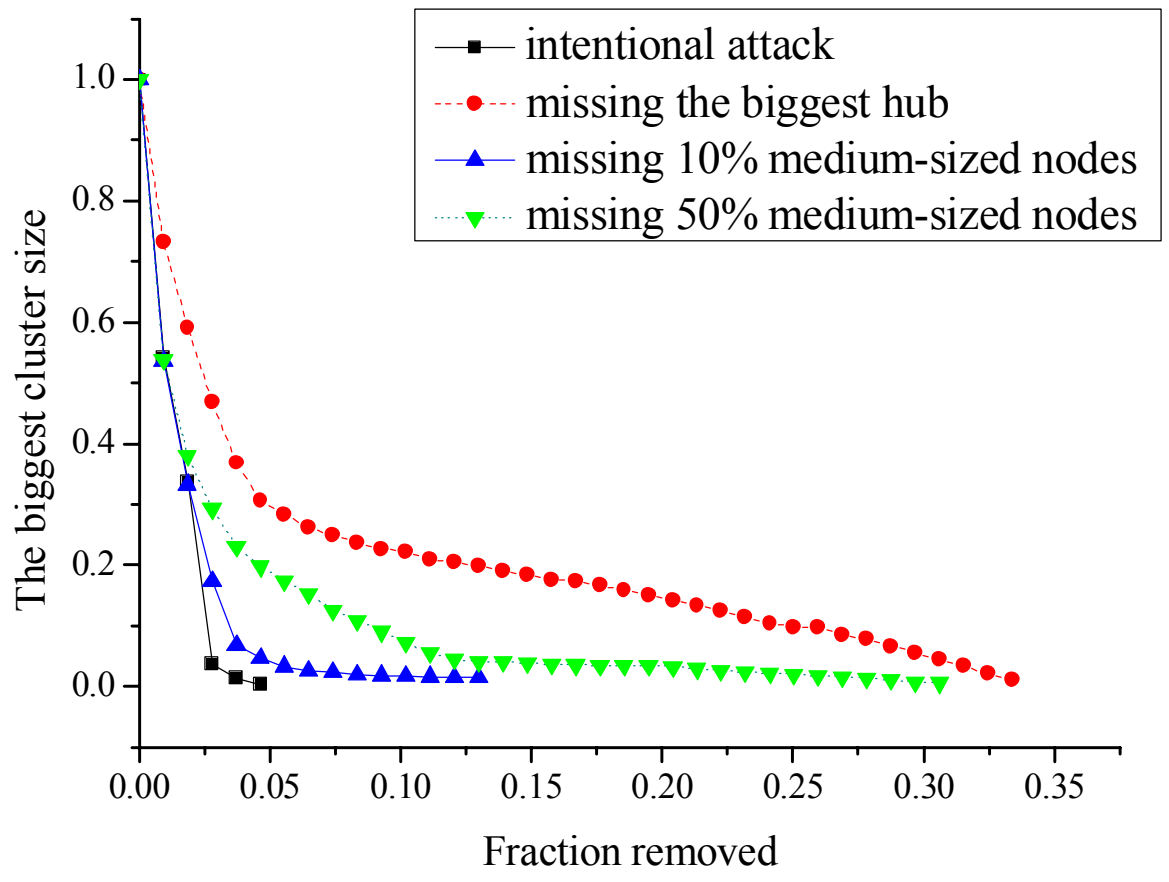

(b) 6470-node Internet model

Fig. 2. Robustness of scale-free networks under intentional attacks that miss (i) the biggest hub, and (ii) 10% and 50% of medium-sized nodes, respectively.

Another interesting observation is that in the BA model, missing 50% medium-sized nodes actually leads to higher threshold of crash than that of missing the biggest hub (Fig. 2(a)), which is obviously not the case in the Internet model (Fig. 2(b)). This again can be explained by the relatively much lower degree of the biggest hub in the BA model: the impact of missing a degree-386 node, significant as it is, is surpassed by that of missing 150 medium-sized nodes with degrees between 12 and 23. The impact of missing the degree-1458 hub in the Internet model, on the other hand, overweighs that of missing 97 nodes with degrees between 10 and 40.

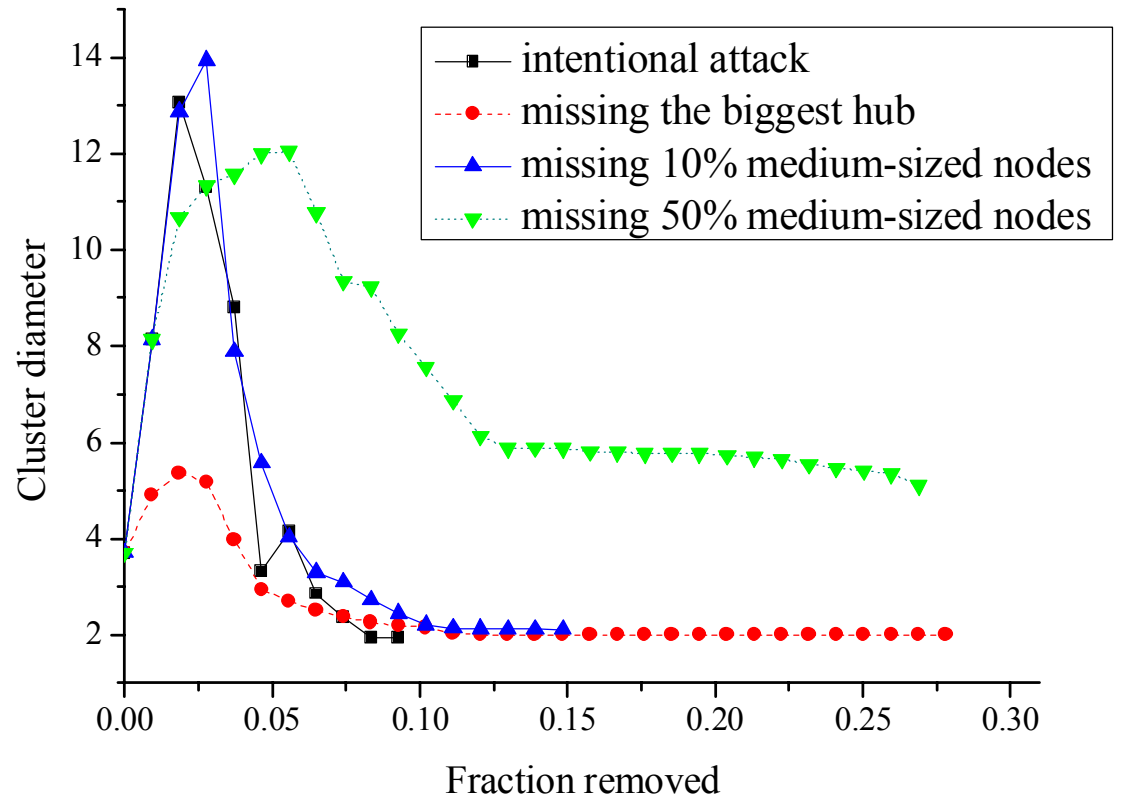

Fig. 3. The cluster diameters of the 6470-node Internet model under intentional attacks.

Finally, the cluster diameters of the Internet model under various types of intentional attacks are shown in Fig. 3. We see that in most cases, the cluster diameters are significantly increased a long time before the network is crashed, making the network highly inefficient in supporting data communications though it is largely still connected. Such impacts can also be degraded when complete knowledge of network topology is not available, especially when a big hub is missed in the attack. As we can see in Fig. 3, when the biggest hub is not taken down, the cluster diameter actually never becomes too large during the whole procedure of attack, until the network is crashed. A similar conclusion holds in the BA model as well.

## III. DISTRIBUTED ATTACKS BASED ON LOCAL INFORMATION

As mentioned before, the distributed attack can only "propagate" through the network step by step, attacking the neighborhood of some nodes that had been crashed before. Furthermore, the target(s) of the next-step attack is selected solely based on the local information of the degrees of the neighborhood nodes.

We evaluate several different distributed attack schemes as follows:

- **Greedy sequential attack:** the attack that chooses the largest-degreed live node adjacent to the node crashed in the *last* step as its next-step target. If no neighborhood live node exists, it randomly selects a live node as its target.
- **Coordinated attack:** the attack that searches through *all* the live nodes adjacent to *any* crashed node and selects among them the largest-degreed one as its next-step target. If no such node exists, it randomly selects a live node to continue the attack.
- **Lower-bounded parallel attack:** instead of taking down only a single node each step, it takes down in each step *all* those live nodes that are (i) adjacent to the crashed nodes, and (ii) with nodal degrees higher than a preset threshold. If no such node exists, the attack stops.

Fig. 4 compares the efficiencies of the greedy sequential attack and the coordinated attack, respectively. It is shown that the greedy sequential attack is not very effective in bringing down the whole network, whereas the coordinated attack, though based on local information only, achieves almost the same threshold of crash as that of the accurate intentional attack. This observation can be understood as follows: we have to take down a lot of low-degreed nodes in the greedy sequential attack since in many steps there are no big hubs adjacent to the just-crashed node. In the coordinated attack, each step we can choose the next-step target from a much larger number of nodes. Consequently we seldom need to attack a low-degreed node. The main drawback of the coordinated attack, however, is the rather complicated collaborations between all the attacking frontiers in order to find the best next-step target.

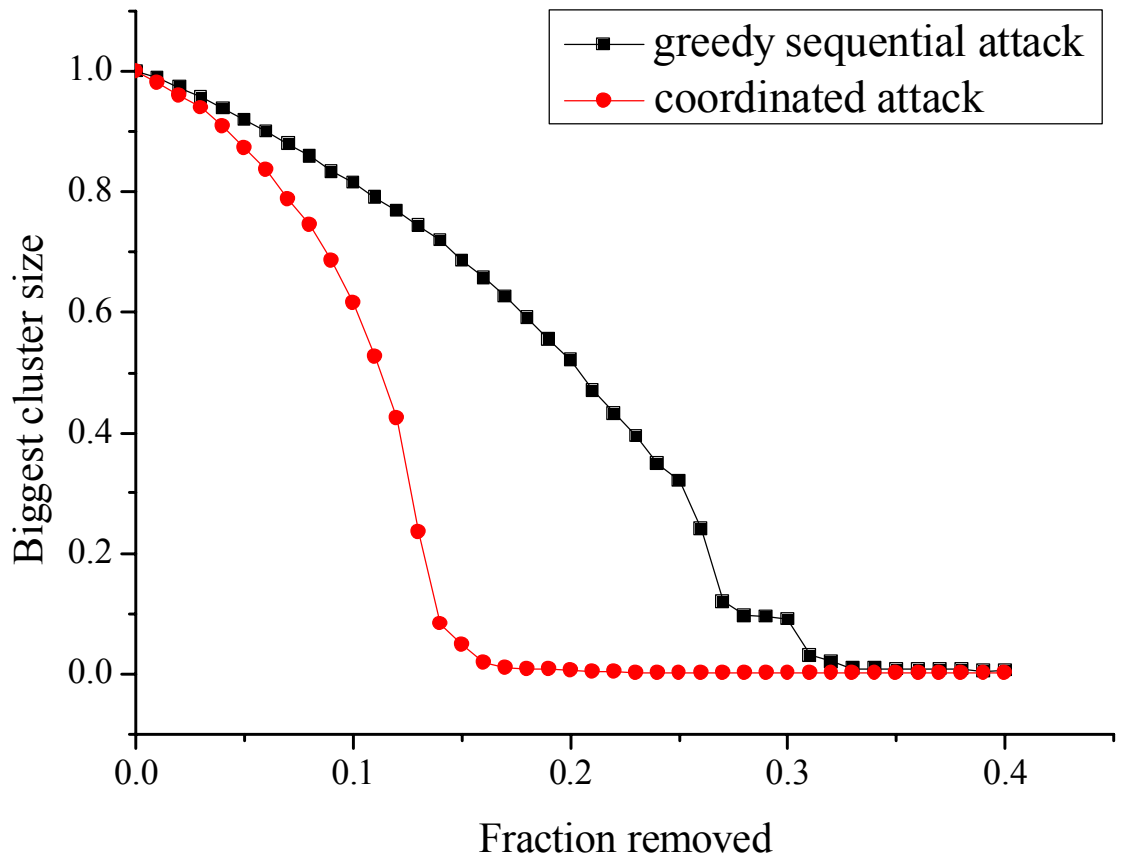

(a) BA model

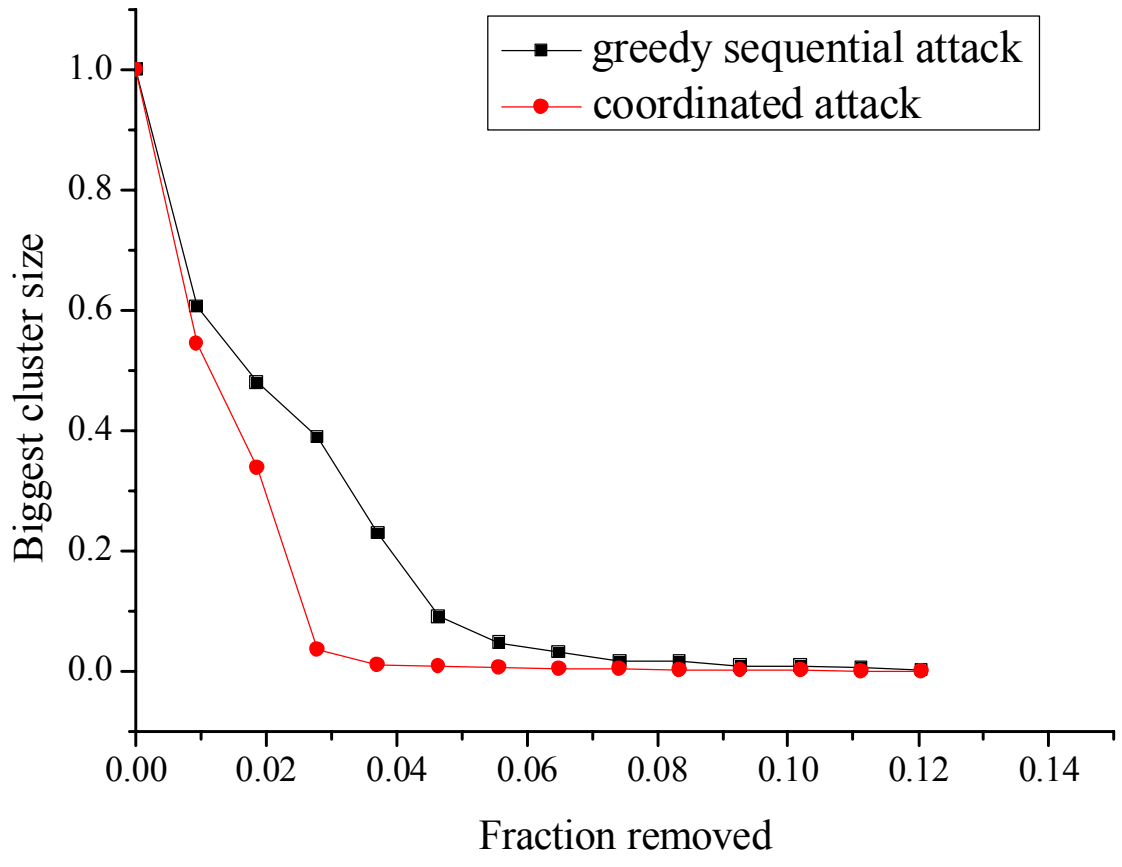

(b) 6470-node Internet model

Fig. 4. The efficiencies of the greedy sequential attack and the coordinated attack, respectively.

To eliminate the need for having complicated collaborations and information exchanges between all the attacking frontiers, we proposed the lower-bounded parallel attack, which arguably may better resemble the real-world attacks as well. To avoid the big burden of taking down a large number of low-degreed nodes, we set up nodal-degree lower bounds at 4 and 10 respectively. A live node adjacent to a crashed node will not be taken down unless its current degree is strictly higher than the threshold.

The simulation results are reported in Fig. 5. In the Internet model, setting the nodal-degree lower bound at 10 makes the distributed attack almost as effective as the accurate intentional attack. In fact, the threshold of crash is only slightly increased from 2.7% to 2.9%. If we set the lower bound at 4, more nodes have to be taken down before the network goes crashed (The threshold of crash becomes 4.8%.). On the other hand, fewer steps of attacks are needed since more nodes are taken down in each step. In the BA model, the observations are different: when we set the nodal-degree lower bound at 4, the threshold of crash is close to that of the accurate intentional attack (16% vs. 14%). When the threshold is set at 10, however, the attack will be terminated after 17 steps when the degrees of all the live nodes adjacent to the attacking frontiers are lower than the threshold. At that moment, the network is largely still connected.

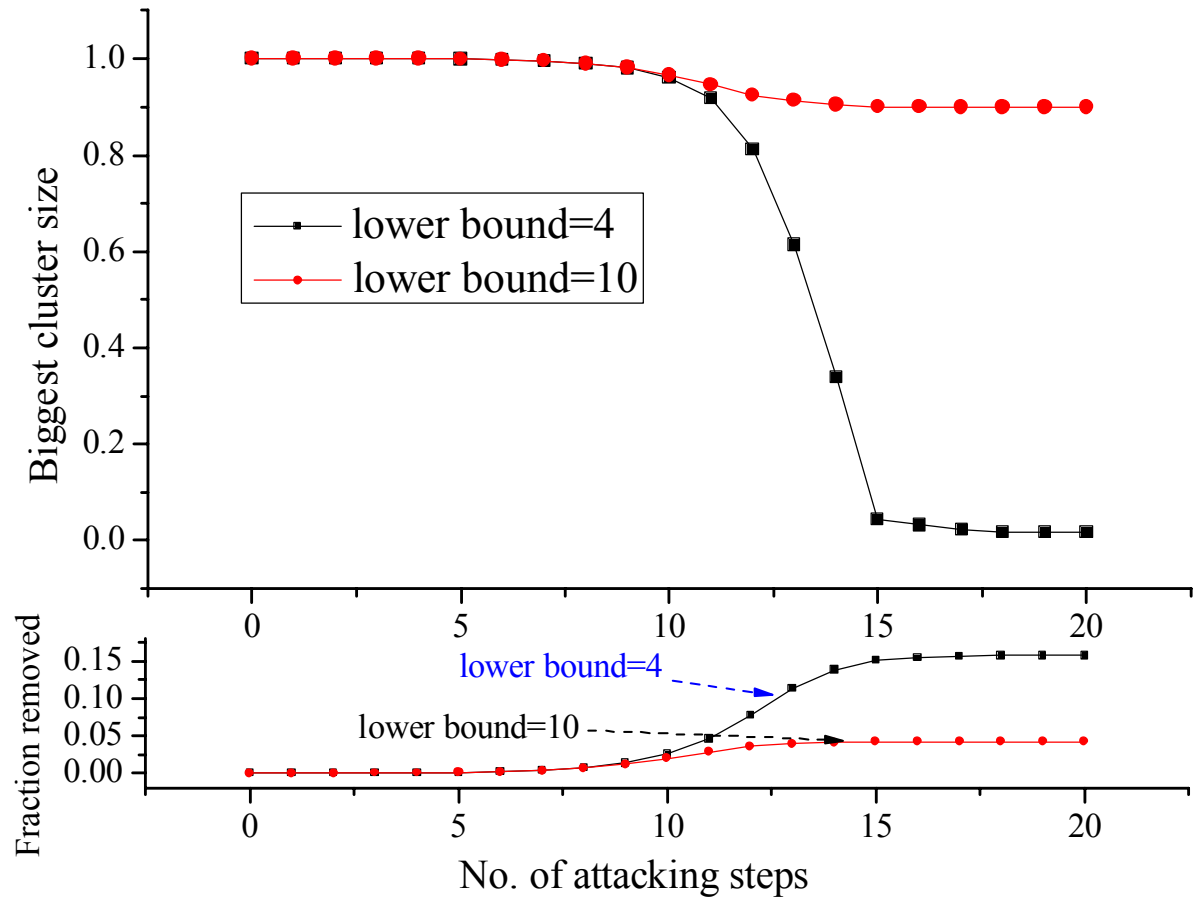

(a) BA model

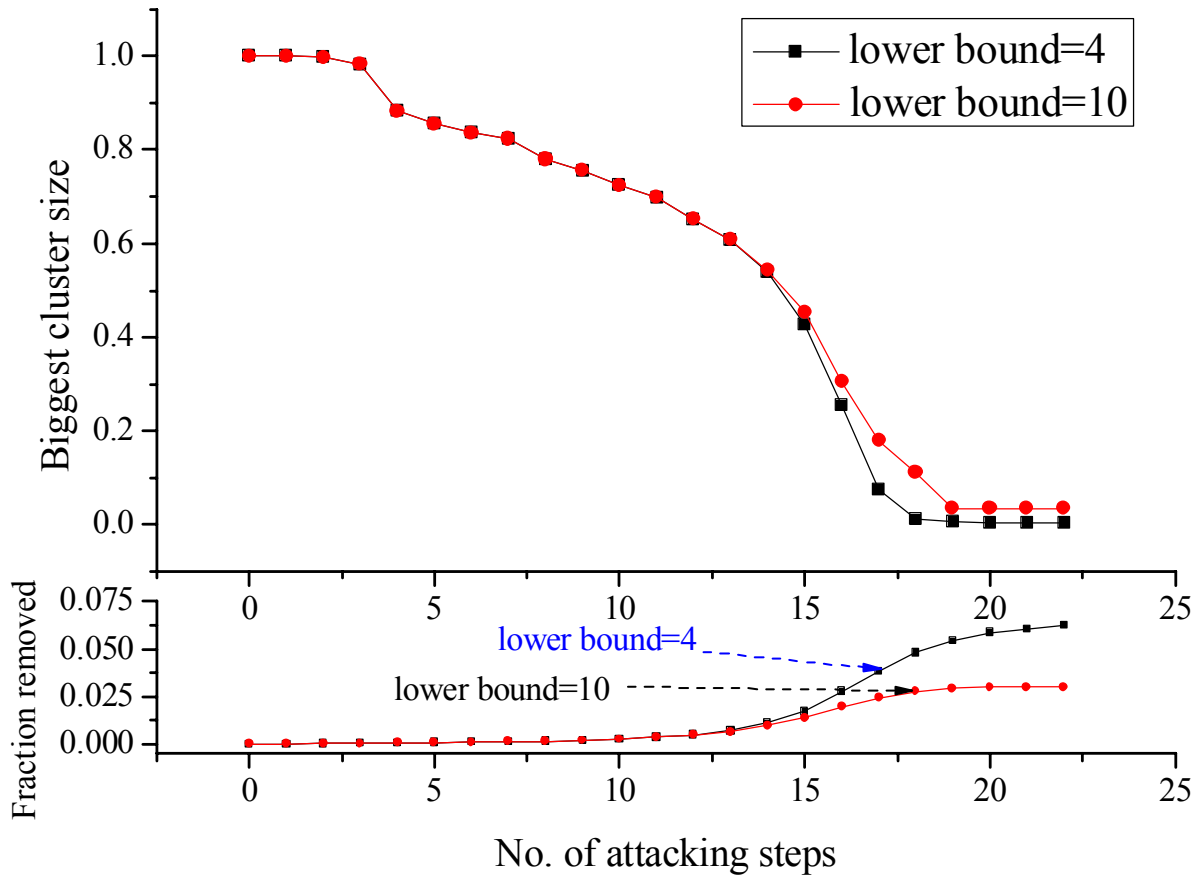

(b) 6470-node Internet model

Fig. 5. The efficiencies of the lower-bounded parallel attack with nodal-degree lower bounds set at 4 and 10, respectively.

The above observations can be explained as follows: the large-degreed nodes in most complex networks form up a *connected* sub-network. Therefore we could attack and take down this sub-network virtually without going through any low-degreed nodes. More significantly, to take down these large-degreed nodes, we actually do not need global information of network topology at all: local information-based attacks can penetrate through the whole sub-network thanks to the connectivity of it. To protect the communication systems from being crashed by such distributed attacks, it is crucially important to identify and stop the attack at early stage. As we can see in Fig. 5, the networks' decomposition is slow at the early stage and then is tremendously speeded up after a certain number of steps. In addition, we argue that it is important for the offending side to select proper threshold values in such attacks: a too-low threshold makes the attack unnecessarily take down a large number of low-degree nodes, slowing down the attack and giving the defending side a better chance to react. A too-high threshold, on the other hand, makes the attack unable to crash the network, as shown in Fig. 5(a).

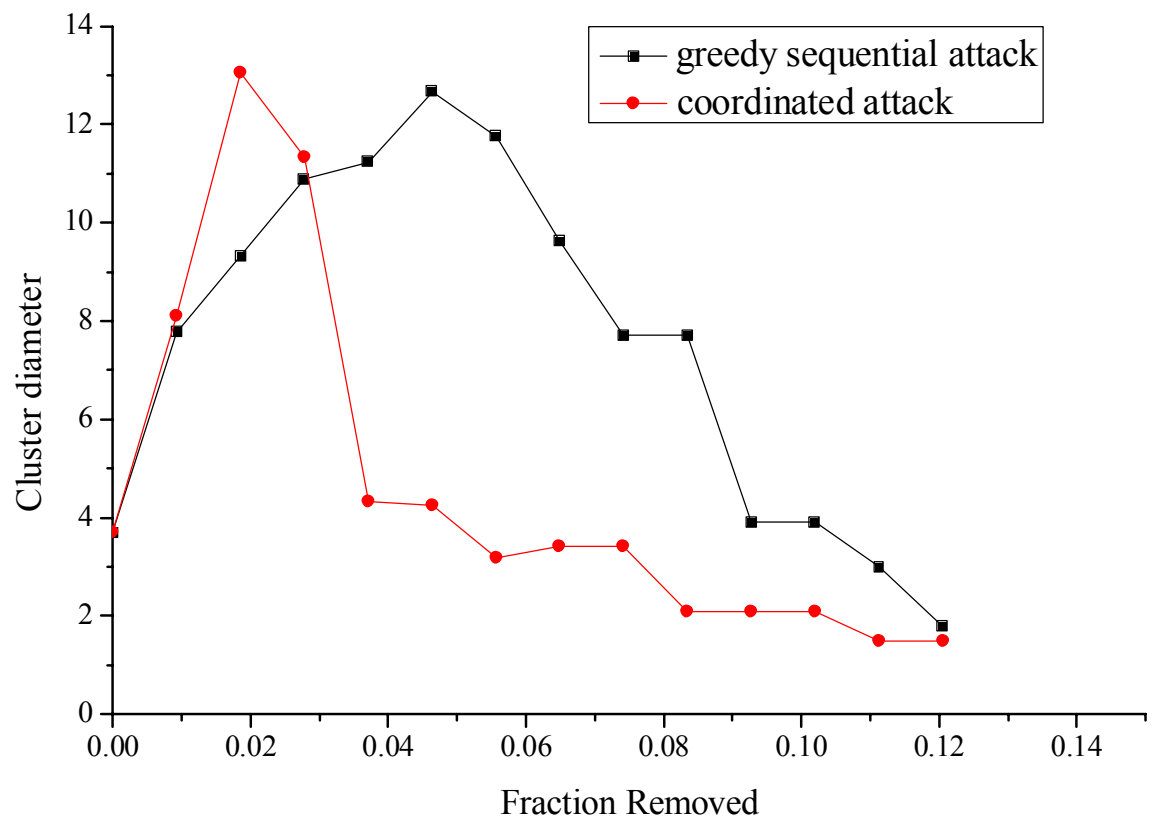

(a) Greedy sequential attack and coordinated attack

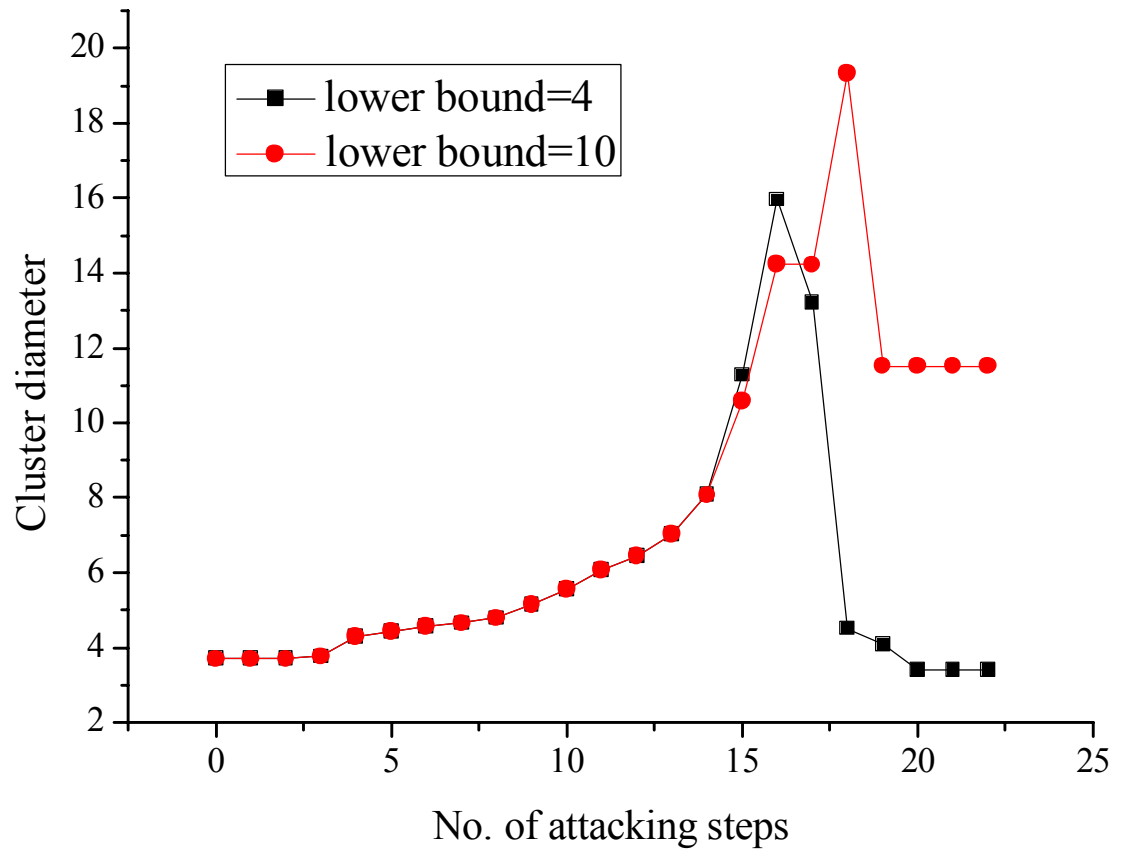

(b) Lower-bounded parallel attack

Fig. 6. The cluster diameters of the 6470-node Internet model under different distributed attacks.

Lastly, we examine the cluster diameters of the Internet model under distributed attacks. We observe in Fig. 6(a) that the coordinated attack quickly makes cluster diameter quite large; and shortly after that, the network is totally crashed. The greedy sequential attack, though not very effective in crashing the whole network, quickly makes the cluster diameter quite long (and hence makes the network no longer efficient in supporting data transmission) as well. Another interesting observation is that, under the greedy sequential attack, even after the network has been largely crashed with 9% network nodes removed, the cluster diameter remains close to 8, indicating the existence of some sparsely-connected clusters (which are of small sizes yet large diameters). In the lower-bounded parallel attacks, as we see in Fig. 6(b), the cluster diameter increases slowly at the beginning and then jumps high very quickly. Also we see that, when we set the lower bound at 10, the cluster diameter remains a constant value 12 after 19 steps of attacks. This large diameter belongs to a sparsely-connected 217-node cluster in which there is no node with a nodal degree higher than 10.

## IV. Conclusion

In this paper, we examined the robustness of complex communication networks under intentional attacks. We considered different cases where the attacks are based on incomplete/inaccurate network-topology information and local information respectively. We found that incomplete information may degrade the efficiency of intentional attack significantly, especially if a big hub is missed. On the other hand, local-information based distributed attacks can be highly effective, sometimes almost as efficient as the accurate global information-based attack. Such insights, as we believe, will be the helpful for developing effective attacking/protecting strategies in the future complex communication systems.